\documentclass[preprint,3p,times]{elsarticle}

\usepackage[T1]{fontenc} 
\usepackage[utf8]{inputenc} 
\usepackage{times}
\usepackage{graphicx}
\usepackage{ifthen}
\usepackage{xspace}
\usepackage{alltt}
\usepackage{latexsym}
\usepackage[scaled=0.86]{helvet}
\usepackage{url}
\usepackage{amssymb}
\usepackage{amsfonts}
\usepackage{amsmath}
\usepackage{stmaryrd}
\usepackage{enumerate}
\usepackage[pdftex,colorlinks=true,pdfstartview=FitV,linkcolor=blue,citecolor=blue,urlcolor=blue]{hyperref}
\usepackage{xspace}
\newboolean{showcomments}
\setboolean{showcomments}{false}
\ifthenelse{\boolean{showcomments}}
  {\newcommand{\bnote}[2]{
	\fbox{\bfseries\sffamily\scriptsize#1}
    {\sf\small$\blacktriangleright$\textit{#2}$\blacktriangleleft$}
   }
   
  }
  {\newcommand{\bnote}[2]{}
   
  }

\newcommand{\sd}[1]{\bnote{Stef}{#1}}

\newcommand{\mmp}[1]{\bnote{Mariano}{#1}}
\newcommand{\noury}[1]{\bnote{Noury}{#1}}

\graphicspath{{figures/}}

\newcommand{\commented}[1]{}
\newcommand{\secref}[1]{Section \ref{sec:#1}}
\newcommand{\seclabel}[1]{\label{sec:#1}}

\newcommand{\stSelector}[1]{{\textsf{\##1}}\xspace}

\newcommand{\eg}{\emph{e.g.,}\xspace}
\newcommand{\ie}{\emph{i.e.,}\xspace}
\newcommand{\ct}[1]{{\textsf{#1}}\xspace}

\newenvironment{code}
    {\begin{alltt}\sffamily}
    {\end{alltt}\normalsize}

\newcommand{\defaultScale}{0.55}

\usepackage{url}            
\makeatletter
\def\url@leostyle{%
  \@ifundefined{selectfont}{\def\UrlFont{\sf}}{\def\UrlFont{\small\sffamily}}}
\makeatother
 \def\signed #1{{\leavevmode\unskip\nobreak\hfil\penalty50\hskip2em
  \hbox{}\nobreak\hfil(#1)%
  \parfillskip=0pt \finalhyphendemerits=0 \endgraf}}

\newsavebox\mybox
\newenvironment{aquote}[1]
  {\savebox\mybox{#1}\begin{quote}\itshape}
  {\signed{\usebox\mybox}\end{quote}}

\usepackage[table]{xcolor}
\usepackage{tabularx}
\rowcolors{1}{black!10}{}

\newboolean{nextRetrait}
\setboolean{nextRetrait}{false}

\newcommand{\retrait}
    {\ifthenelse{\boolean{nextRetrait}}%
       {\hspace{-3cm}}%
       {}}%

\usepackage[normalem]{ulem} 
\usepackage{xcolor}
\newboolean{showedits}
\setboolean{showedits}{true} 
\ifthenelse{\boolean{showedits}}
{
  \newcommand{\del}[1]{\textcolor{red}{\sout{#1}}} 
}{
  \newcommand{\del}[1]{} 
  
}

\newcommand{\dnu}[1]{\stSelector{doesNotUnderstand:}}
\newcommand{\cannot}[1]{\stSelector{cannotInterpret:}}

\begin{document}


\newcommand{\affINRIA}{$^{1}$}
\newcommand{\affEMD}{$^{2}$}

\begin{frontmatter}

\title{Ghost: A Uniform and General-Purpose Proxy Implementation}
\tnotetext[t1]{This work was supported by Ministry of Higher Education and Research, Nord-Pas de Calais Regional Council and FEDER through the 'Contrat de Projets Etat Region (CPER) 2007-2013'.}

\author{Mariano Martinez Peck\corref{cor1}\fnref{fn1,fn2}}
\ead{marianopeck@gmail.com}
\author{Noury Bouraqadi\fnref{fn2}}
\ead{noury.bouraqadi@mines-douai.fr}
\author{St\'ephane Ducasse\fnref{fn1}}
\ead{stephane.ducasse@inria.fr}
\author{Luc Fabresse\fnref{fn2}}
\ead{luc.fabresse@mines-douai.fr}
\author{Marcus Denker\fnref{fn1}}
\ead{marcus.denker@inria.fr}

\cortext[cor1]{Corresponding author}
\fntext[fn1]{RMoD Project-Team, Inria Lille--Nord Europe / Universit\'e de Lille 1.}
\fntext[fn2]{Universit\'e Lille Nord de France, Ecole des Mines de Douai.}

\begin{abstract}

A proxy object is a surrogate or placeholder that controls access to another target object.  Proxy objects are a widely used solution for different scenarios such as remote method invocation, future objects, behavioral reflection, object databases, inter-languages communications and bindings, access control, lazy or parallel evaluation, security, among others.

Most proxy implementations support proxies for regular objects but are unable to create proxies for objects with an important role in the runtime infrastructure such as classes or methods. Proxies can be complex to install, they can have a significant overhead, they can be limited to certain kind of classes, etc. Moreover, proxy implementations are often not stratified and they do not have a clear separation between proxies (the objects intercepting messages) and handlers (the objects handling interceptions).

In this paper, we present Ghost: a uniform and general-purpose proxy implementation for the Pharo programming language. Ghost provides low memory consuming proxies for regular objects as well as for classes and methods.

When a proxy takes the place of a class, it intercepts both the messages received by the class and the lookup of methods for messages received by its instances. Similarly, if a proxy takes the place of a method, then the method execution is intercepted too.

\end{abstract}

\end{frontmatter}

\section{Introduction}
\seclabel{intro}

A proxy object is a surrogate or placeholder that controls access to another target object. A large number of scenarios and applications have embraced and used the Proxy Design Pattern  \cite{Gamm93b,Eugs06a}. Proxy objects are a widely used solution for different scenarios such as remote method invocation \cite{Shapi86,nuno02}, distributed systems \cite{Benn87a,McCu87a}, future objects \cite{Prat04a}, behavioral reflection \cite{Duca99a,Welch99}, wrappers \cite{Bran98a}, object databases \cite{proxiesDatabase}, inter-languages communications and bindings, access control and read-only execution \cite{Arna10a}, lazy or parallel evaluation, middlewares like CORBA \cite{Wang01,Koster00,Hasso05}, encapsulators \cite{Pasc86a},  security \cite{Vanc10a},  memory management and object swapping  \cite{Mart11b,Mart11c}, among others.

Most proxy implementations support proxies for instances of common classes only. Some of them, \eg Java Dynamic Proxies \cite{Eugs06a}, even need that at creation time the user provides a list of \emph{Java interfaces} for capturing the appropriate messages.

In the context of class-based dynamic object-oriented languages that reify those entities with an important role in the runtime infrastructure such as classes or methods with first-class objects, proxies can be created only for regular objects. Creating uniform proxies for other entities such as classes or methods has not been considered.
In existing work, it is impossible for a proxy to take the place of a class and a method and still be able to intercept messages and perform operations such as logging, security, remote class interaction, etc.

For example, imagine a virtual memory for dynamic languages whose goal is to use less memory by only leaving in primary memory what is needed and used, swapping out the unused objects to secondary memory \cite{Kaeh86a,Bond07b,Mart11b,Mart11c}. To achieve this, the system replaces the original (unused) objects with proxies. When one of the proxies intercepts a message, the original object is brought back into primary memory. In this system, the original objects can be instances of common classes but they can also be methods, classes, method context themselves, etc. Therefore, a proxy implementation must deal with all kind of objects including classes and methods because this weakness strongly limits the application of proxies.

Another property of proxy implementations is memory footprint. As any other object, proxies occupy memory and there are scenarios \eg the previously mentioned object graph swapper, where the number of proxies and their memory footprint becomes a problem.

Traditional implementations in dynamic languages such as Smalltalk are based on error handling \cite{Pasc86a}. This results in non stratified proxies meaning that not all messages can be trapped leading to severe limits. Not being able to intercept messages is a problem because those messages will be directly executed by the proxy instead of being intercepted. This can lead to different execution paths in the code, errors or even a VM crash.

Traditionally, proxies not only intercept messages, but they also decide what to do with the interceptions. We argue that these are two different responsibilities that should be separated. Proxies should only intercept, which is a generic operation that can be reused in different contexts. Processing interceptions is application-dependent. It should be the responsibility of another object that we call \emph{handler}.

In this paper, we present Ghost: a uniform and general-purpose proxy implementation for the Pharo programming language \cite{Blac09a}. Ghost provides low memory consuming proxies for regular objects, classes and methods. It is possible to create a proxy that takes the place of a class or a method and that intercepts messages without breaking the system. When a proxy takes the place of a class, it intercepts both the messages received by the class and the lookup of methods for messages received by instances. Similarly, when a proxy takes the place of a method, then the method execution is intercepted too.  Last, Ghost supports \emph{controlled stratification}: developers decide which message should be understood by the proxy and which should be intercepted and transmitted for processing to the handler.

This paper presents an extension of our previous work on proxies published as a workshop paper \cite{Mart11a}. We have removed the section ``Ghost model'' and written new sections and subsections: ``Extending and Adapting Proxies and Handlers'', ``Intercepting Everything or Not?'', ``Messages to be Answered by the Handler'', ``Special Messages and Operations'', ``Case Studies'', ``Discussion''. Apart from that, several sections were extensively extended, rewritten or improved.

The contributions of this paper are:

\begin{itemize}
\item Describe and explain the common proxy implementation in dynamic languages and, specially, in Pharo.

\item Define a set of criteria to evaluate and compare proxies implementations.

\item Present Ghost: a new proxy implementation which solves most of the problems with uniform proxies.

\item Validate our solution using the defined criteria as well as the speed and memory consumption of Ghost.

\end{itemize}

The remainder of the paper is structured as follows: \secref{proxySurvey} defines and unifies the vocabulary and roles used throughout the paper and presents the list of criteria used to compare different proxy implementations. \secref{commonProxy} describes the typical proxy implementation and it presents the problem by evaluating it against the previously defined criteria. An introduction to Pharo reflective model and its provided hooks is explained by \secref{smalltalkIntroAndSupport}. \secref{ghostImpl} introduces and discusses Ghost proxies and shows how the framework works. \secref{proxiesClassesAndMethods} explains how Ghost is able to proxify methods and classes. Certain messages and operations that need special care when using proxies is analyzed in \secref{specialMessages}. \secref{criteriaGhost} provides an evaluation of Ghost based on the defined criteria. \secref{discussion} discusses Ghost model generality.  Real case studies of Ghost are presented in \secref{caseStudies}. Finally, in \secref{relatedWorks}, related work is presented before concluding in  \secref{conclusion}.

\section{Proxy Evaluation Criteria}
\seclabel{proxySurvey}

\subsection{Vocabulary and Roles}
\begin{figure}[ht]
\begin{center}
\includegraphics[width=5cm]{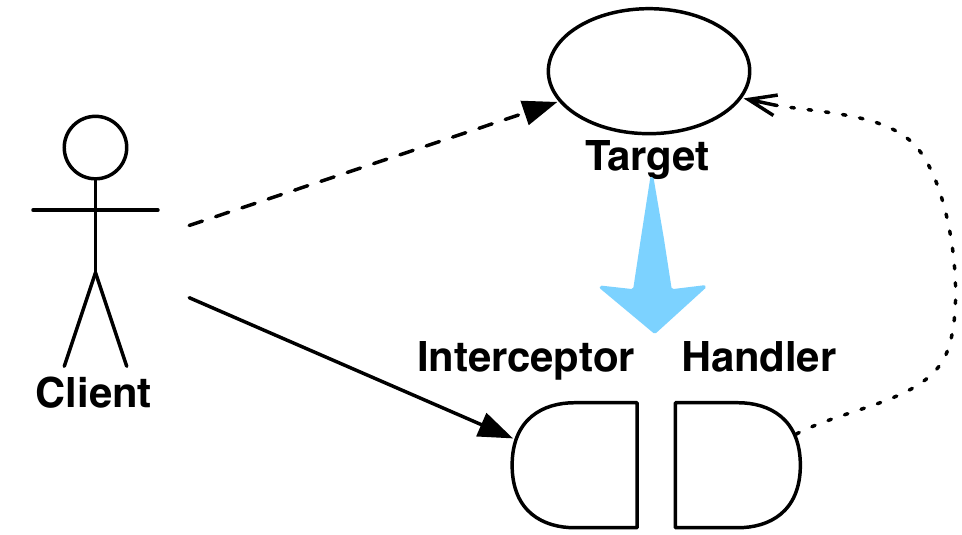}
\caption{Roles in Proxy.\label{role}}
\end{center}
\end{figure}

For sake of clarity, we define here the vocabulary used throughout this paper and we highlight the roles that objects are playing when using proxies (see Figure~\ref{role}).

\begin{description}
\item{\emph{Target.}} It is the original object that we want to \emph{proxify}.
\item{\emph{Client.}}  It is an object which uses or holds a reference to a target object.
\item{\emph{Interceptor.}} It is an object whose responsibility is to \emph{intercept} messages that are sent to it. It may intercept some messages or all of them.
\item{\emph{Handler.}} The handler is responsible of \emph{handling} messages caught by the interceptor. By \emph{handling} we refer to whatever the user of the framework wants to do with the interceptions, \eg logging, forwarding the messages to the target, control access, etc.
\end{description}
One implementation can use the same object for taking the roles of interceptor and handler, \ie the proxy plays both roles. In another solution, such roles can be achieved by different objects. With this approach, the proxy usually takes the role of interceptor.

\subsection{Proxies Implementation Criteria}
\label{sec:criteria}

From the implementation point of view, there are criteria that have to be taken into account to compare and characterize a particular solution \cite{Duca99a,Vanc10a}:

\paragraph{\textbf{Stratification}} Most solutions to implement proxies are based on dedicated messages such as \ct{doesNotUnderstand:}. The problem with approaches that reserve a set of messages for the proxy implementation is that there is a clash between the API of the proxified object and the proxy implementation.

To address such problem, some solutions proposed stratification \cite{Vanc10a}. Stratification means that there is a clear separation between the proxy support and application functionalities.  In a fully stratified proxy, all messages received by a proxy should be intercepted and transmitted for processing to a handler. The proxy API should not pollute the application's namespace. Besides, having this stratification is important to achieve security and to fully support transparency of proxified objects for the end-programmers \cite{Brac04b}.

Stratification highlights two responsibilities in a proxy toolbox:  (1) trapping or intercepting messages (interceptor role) and (2) managing interceptions (handler role), \ie performing actions once messages are intercepted. In a stratified proxy framework, the first responsibility is covered by a proxy itself and the second one by a handler.

\paragraph{\textbf{Interception levels}} There are the following possibilities:
\begin{itemize}
\item Intercept \emph{all} messages, even those not defined in the object API \eg inherited from superclasses.
\item Intercept all messages excluding a list of messages defined by the user.
\item Intercept all messages excluding some messages imposed by the proxy toolbox \eg inherited methods if we are using a solution based on error handling such as using the \ct{doesNotUnderstand:} message.

\end{itemize}

With the last option, the developer has no control over messages that are not intercepted and hence performed by the proxy itself.  This can be a problem because it is impossible to distinguish messages sent to the proxy from the ones that should be trapped. For example, when a proxy is asked its class, it must answer not its own class but the class of the target object. Otherwise, this can cause errors difficult to manage.

\paragraph{\textbf{Object replacement}}
Replacement is making client objects refer to the proxy instead of the target. Two cases exist:

\begin{enumerate}
\item Often, the target is an existing object with other objects referencing it. The target needs to be \emph{replaced} by a proxy,  \ie all objects in the system which have a reference on the target should be updated so that they point to the proxy instead. For instance, for a virtual memory management, we need to swap out unused objects and to replace them with proxies. We refer to this functionally as \emph{object replacement}.

\item In the other case, the proxy is just created and it does not replace another already existing object. For example, when doing a query with a database driver, it can create proxies to perform lazy loading on some parts of the graphs. As soon as a proxy receives a message, the database driver loads the rest of the graph. Another example is remote method invocation where targets are located in a different memory space from the clients' one. This means that, in the client memory space, we have proxies that can forward messages and interact with the real objects in the other memory space.
\end{enumerate}

Object replacement is not a feature implemented by the proxy toolbox itself. However, since it allows users to proxify objects, the proxy library must support the case where objects are replaced by proxies. This is a challenge because there are objects that in order to be correctly replaced they need special proxies.

\paragraph{\textbf{Uniformity}} We refer to the ability of creating a proxy for any kind of object (regular object, method, class, block, process, etc) and replacing the object with it. Most proxy implementations support proxies only for regular objects \ie proxies cannot replace a class, a method, a process, etc., without breaking the system. Certain particular objects like \ct{nil}, \ct{true} and \ct{false} cannot be proxified either.

This is an important criterion since there are scenarios where being able to create proxies for any runtime entity is mandatory. As described in \secref{caseStudies} an example is the mentioned virtual memory  which replaces all type of unused objects with proxies

\paragraph{\textbf{Transparency}} A proxy is fully transparent if clients are unaffected whether they refer to a proxy or the target. No matter what message the client sends to the proxy, it should answer the same as if it were the target object.

One of the typical problems related to transparency is the identity issue when the proxy and the target are located in the same memory space.  Given that different objects have different identities, a proxy's identity is different from the target's identity.  The expression \ct{proxy == target} will answer \ct{false} revealing the existence of the proxy.
This can be temporary hidden if there is object replacement between the target object and the proxy. When we replace all references to the target with references to the proxy, clients will only see the proxy.
However, this "illusion" will be broken as soon as the target provides its own reference (\ct{self}) as an answer to a message.

Another common problem is asking a proxy the class or type since, most of the times, the proxy answers its own type or class instead of the one of the target. The same happens if there is special syntax or operators in the language such as ``+'', ``/'',  ``='', ``>'' \cite{Aust11a}. To have the most transparent  proxy possible, these situations should be handled in a way  which allows the proxy to behave like the target.

\paragraph{\textbf{Efficiency}}  The proxy toolbox must be efficient from the performance and memory usage points of view. In addition, we can distinguish between installation performance and runtime performance. For example, for installation, it is commonly evaluated if a proxy installation involves extra overhead like recompiling.

Depending on the usage, the memory footprint of the proxies can be substantial.  The space analysis should consider not only the size in memory of the proxies, but also how many objects are needed per target: it can be either only one proxy instance or a proxy instance and a handler instance.

\paragraph{\textbf{Ease of debugging}}
It is difficult to test and debug in the presence of proxies because debuggers or test frameworks usually send messages to the objects present in the current stack. These messages include, for example, printing an object, accessing its instance variables, etc. When the proxy receives any of these messages it may intercept such message, making debugging more complicated.

\paragraph{Proxy Toolbox and Implementation}

\mmp{remove this paragraph?}

The proxy toolbox and its implementation can also raise some specific concerns:
\begin{description}

\item[\emph{Implementation complexity.}]   This criterion measures how difficult it is to implement a solution. Given a fixed set of functionalities, a simpler implementation is better.

\item[\emph{Constraints.}]  The toolbox may require, for example, that the target implements certain interface or inherits from a specific class. It is important that the user of the proxy toolbox can easily extend or adapt it to his own needs.

\item[\emph{Portability.}]  A proxy implementation can depend on specific entry points of the virtual machine or on certain features provided by the language.
\end{description}

\section{Common Proxy Implementations}
\seclabel{commonProxy}

Although there are different proxy implementations and solutions, there is one that is the most common among dynamic programming languages. This implementation is based on error raising and the resulting error handling \cite{Duca99a,Blac09a}. We briefly describe it and show that it fails to fulfill important requirements.

\subsection{Typical Proxy Implementation}

In dynamic languages, the type of the object receiving a message is resolved at runtime. When an unknown message is sent to an object, an error exception is thrown. The basic idea is to create objects that raise errors for all the possible messages (or a subset) and customize the error handling process.

In Pharo, for instance, the virtual machine sends the message \ct{doesNotUnderstand:} to the object that receives a message that does not match any method. To avoid infinite recursion, all objects must understand the message \ct{doesNotUnderstand:}. That is the reason why such method is implemented in the class \ct{Object}, the root of the hierarchy chain. The default implementation throws a \ct{MessageNotUnderstood} exception. Similar mechanisms exist in dynamic languages like Ruby, Python, Objective-C and Perl.

Since \ct{doesNotUnderstand:} is a normal method, it can be overwritten in subclasses. Hence, if we can have a minimal object and we override the \ct{doesNotUnderstand:} method to do something special (like forwarding messages to a target object), then we have a possible proxy implementation. This technique has been used for a long time \cite{McCu87a,Pasc86a} and it is the most common proxy implementation. Readers knowing this topic can directly jump to \secref{commonProxyEvaluation}. Most dynamic languages provide a mechanism for handling messages that are not understood as shown in \secref{relatedWorks}.

\paragraph{Obtaining a minimal object}  A minimal object is one which understands none or only a few methods \cite{Duca99a}. In some programming languages, the root class of the hierarchy chain (usually called \ct{Object}) already contains several methods.
In Pharo, \ct{Object} inherits from a superclass called \ct{ProtoObject} which inherits from \ct{nil}. \ct{ProtoObject} understands a few messages\footnote{ProtoObject has 25 methods in PharoCore 1.4.}: the minimal messages that are needed by the system. Here is a simple Proxy implementation.

\begin{figure}[ht]
\begin{minipage}[c]{7.5cm}
\begin{code}{}
ProtoObject subclass: #Proxy
      instanceVariableNames: 'targetObject'
      classVariableNames: ''
      poolDictionaries: ''
      category: 'Proxies'
\end{code}
\end{minipage}
\hfill
\begin{minipage}[c]{7.8cm}
      \begin{code}{}
Proxy >> doesNotUnderstand: aMessage
    |result|
    ...      "Some application-specific code"
    result := aMessage sendTo: targetObject.
    ...      "Other application-specific code"
      ^result
\end{code}
\end{minipage}
\caption{Naive proxy implementation based in minimal object and handling not understood methods in Pharo.}
\end{figure}

\paragraph{Handling not understood methods}  Common behaviors of proxies include logging before and after the method, forwarding the message to a target object, validating some access control, etc. If needed, it is valid to issue a super send to access the default \ct{doesNotUnderstand:} behavior.

To be able to forward a message to an object, the virtual machine usually reifies the message. In Pharo, the argument of the \ct{doesNotUnderstand:} message is an instance of the class \ct{Message}. It specifies the method selector, the list of arguments and the lookup class (in normal messages it is the class of the receiver and, for super sends, it is the superclass of the class where the method issuing the super send is implemented). To forward a message to another object, the class \ct{Message} provides the method \ct{sendTo: anotherObject}.

This solution is independent of Pharo. For example, the Pharo's \ct{doesNotUnderstand:} and \ct{sendTo:} are in Ruby \ct{method\_missing} and \ct{send}, in Python \ct{\_\_getattr\_\_} and \ct{getattr}, in Perl \ct{autoload}, in Objective-C \ct{forwardInvocation:}. In \secref{relatedWorks}, we explain some of these examples with more detail.

\subsection{Evaluation}
\seclabel{commonProxyEvaluation}

We now evaluate the common proxy implementation based on the criteria we described in \secref{criteria}.

\paragraph{Stratification}   This solution is unstratified:

\begin{itemize}

\item The method  \ct{doesNotUnderstand:} cannot be trapped like a regular message. Moreover, when such message is sent to a proxy, there is no efficient way to know whether it was because of the regular error handling procedure or because of a proxy trap that needs to be handled.  In other words, the \ct{doesNotUnderstand:} occupies the same namespace as application-level methods \cite{Vanc10a}.

\item There is no separation between proxies and handlers.

\end{itemize}

\paragraph{Interception levels}  It cannot intercept all messages but  \emph{only} those that are not understood. As explained, this generates method name collisions.

\paragraph{Object replacement}  It is usually unsupported by most programming languages. Nevertheless, Smalltalk implementations do support it using pointer swapping operations such as the \ct{become:} primitive. However, with such solution, target references may leak when the target remains in the same memory: the target might provide its own reference as a result of a message. This way the client gets a reference to the target so it can by-pass the proxy..

\paragraph{Uniformity}  There is a severe limit to this implementation since it is not uniform: proxies can only be applied to regular objects. \secref{pharoReflective} explains why certain kinds of objects like classes, methods and other core objects need special handling to be correctly proxified.

\paragraph{Transparency}  This solution is not transparent. Proxies do understand some methods (those from its superclass) generating method name collisions. For instance, if we evaluate \ct{Proxy new pointersTo}\footnote{\ct{pointersTo} is a method implemented in \ct{ProtoObject}.} it answers the references to the proxy instead of intercepting the message and forwarding it to a target. The same happens with the identity comparison or when asking the class.

In Pharo, it is possible not only to subclass from \ct{ProtoObject} but also from \ct{nil} in which case the subclass do not inherit any method. This solves some of the problems, such as the one of method name collisions, but the solution is still not stratified and makes debugging more complicated.

\paragraph{Efficiency} From the speed point of view, this solution is reasonably fast (it is based on two lookups: one for the original message and one for the \ct{doesNotUnderstand:} message) and it has low overhead. In contrast to other technologies, there is no need to recompile the application and the system libraries or to modify their bytecode or to do other changes. Regarding the memory usage, there is no optimization.

\paragraph{Ease of debugging}  The debugger sends messages to the proxy which are not understood and, therefore, intercepted. To be able to debug in presence of proxies, one has to implement all these methods directly in the proxy.  The drawback is that the action of enabling or disabling the debugging facilities means adding or removing methods from the proxy. Instead of implementing the methods in the proxy, we could also have a trait (if the language provides traits or any other composable unit of behavior) with such methods. However, we still need to add the trait to the proxy class when debugging and remove it when we are not.

\paragraph{Implementation complexity} This solution is easy to implement: it just requires to create a subclass of the minimal object and implement the \ct{doesNotUnderstand:} method.

\paragraph{Constraints} This solution is flexible since target objects do not need to implement any interface or method, nor to inherit from specific classes. The user can easily extend or change the purpose of the proxy adapting it to his own needs by just reimplementing the \ct{doesNotUnderstand:}.

\paragraph{Portability} This approach needs just a few requirements that have to be provided by the language and the VM. Moreover, almost all available dynamic languages support these needs by default:  a message like \ct{doesNotUnderstand:}, a minimal object and the possibility to forward a message to another object. Therefore, it is easy to implement this approach in different dynamic languages.

\section{Pharo Support for Proxies}
\seclabel{smalltalkIntroAndSupport}

Before presenting Ghost, we first explain the basis of the Pharo reflective model and the provided hooks that our solution uses.
We show that Pharo provides, out of the box, all the support we need for Ghost's implementation \ie object replacement, interception of methods' execution and interception of all messages.

\subsection{Pharo Reflective Model and VM Overview}
\seclabel{pharoReflective}

\begin{figure}[ht]\centering
     \includegraphics[width=0.6\linewidth]{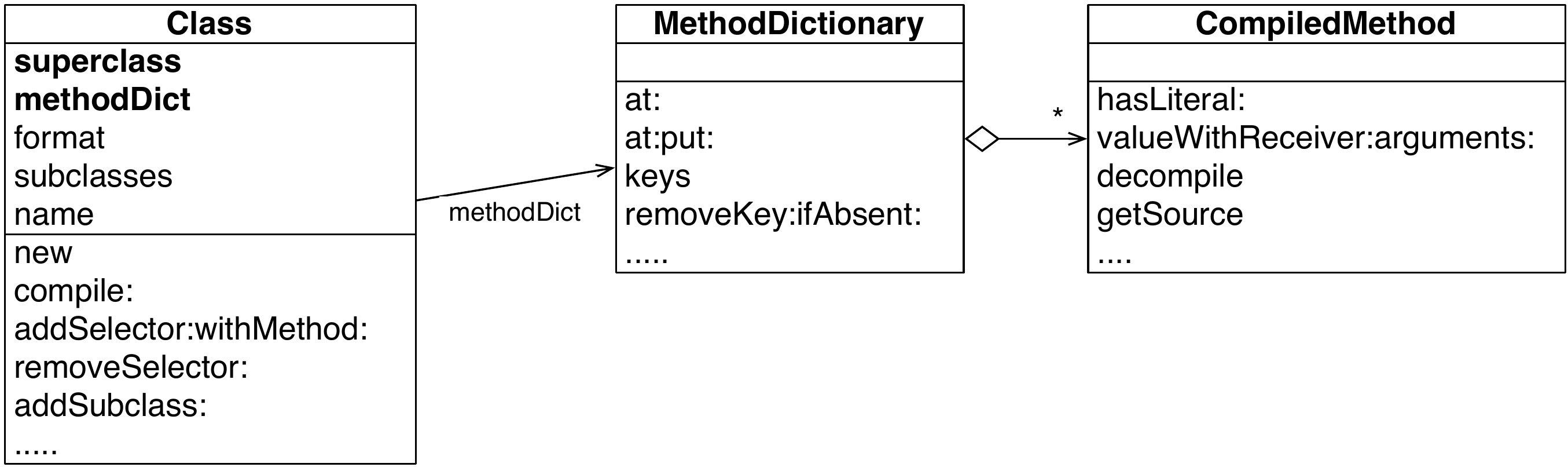}
     \caption{The basic Smalltalk reflective model. Bold instance variables are imposed by virtual machine logic.}
     \label{fig:Metamodel}
\end{figure}

Being a Smalltalk dialect, Pharo inherits the simple and elegant reflective model of Smalltalk-80. There are two important rules \cite{Blac09a}: 1) \emph{Everything is an object}; 2) \emph{Every object is instance of a class}. Since classes are objects and every object is an instance of a class, it follows that classes must also be instances of classes. A class whose instances are classes is called a metaclass.
Figure~\ref{fig:Metamodel} shows a \emph{simplified} reflective model of Smalltalk.

Figure~\ref{fig:Metamodel} shows that a class is defined by a superclass, a method dictionary, an instance format, subclasses, name and a few others. The important point here is that the first two are imposed by the virtual machine\footnote{The VM actually needs three instances variables, the third being the \ct{format}. But, the \ct{format} is accessed only by a few operations \eg instance creation. Since the proxy intercepts all messages including creational ones, the VM will never need to access the \ct{format} while using a proxy.}. The method dictionary is a hash table where keys are the methods' names (called selectors in Smalltalk) and the values are the compiled methods which are instances of \ct{CompiledMethod}.

\subsection{Hooks and Features Provided by Pharo}

The following is a list of the Pharo reflective facilities and hooks that Ghost uses for implementing proxies.

\paragraph{Class with no method dictionary} When an object receives a message and the VM does the method lookup, if the method dictionary of the receiver class (or of any other class in the hierarchy chain) is \ct{nil}, the VM sends the message \ct{cannotInterpret: aMessage} to the receiver.
Contrary to normal messages, the lookup for the method \ct{cannotInterpret:} starts in the \emph{superclass} of the class whose method dictionary was \ct{nil}. Otherwise, there would be an infinite loop. This hook is powerful for proxies because it let us intercept \emph{all} messages that are sent to an object.

\begin{figure}[ht]\centering
     \includegraphics[width=0.7\linewidth]{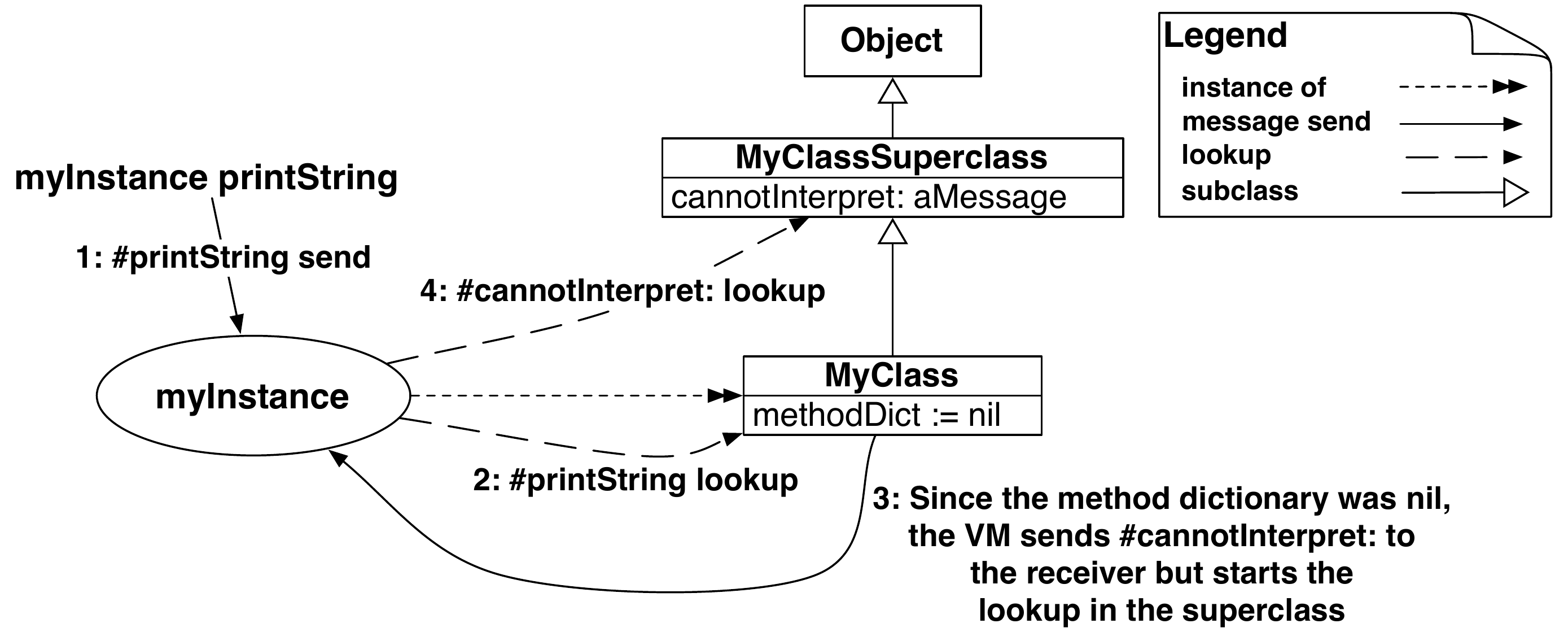}
     \caption{Message handling when a method dictionary is \ct{nil}.\label{fig:cannotInterpret}}
\end{figure}

Figure~\ref{fig:cannotInterpret} depicts the following situation: we get one object called \ct{myInstance}, instance of the class \ct{MyClass} whose method dictionary is \ct{nil}. This class has a superclass called \ct{MyClassSuperclass}. Figure~\ref{fig:cannotInterpret} shows how the mechanism works when sending the message \ct{printString} to \ct{myInstance}. The message \ct{cannotInterpret:} is sent to the receiver (\ct{myInstance}) but starting the lookup in \ct{MyClassSuperclass}.

\paragraph{Objects as methods} This facility allows us to intercept method executions. We can put an object that is not an instance of \ct{CompiledMethod} in a method dictionary. Here is an example:

\begin{code}{}
MyClass methodDict at: #printString put: Proxy new.
MyClass new printString.
\end{code}

When the \ct{printString} message is sent, the VM does the method lookup and finds an entry for \ct{\#printString} in the method dictionary. Since the object associated with the \ct{printString} selector is not a compiled method, the VM sends a special message \ct{run: aSelector with: arguments in: aReceiver} to that object, \ie the one that replaces the method in the method dictionary.

The VM does not impose any shape to objects acting as methods such as having certain amount of instance variables or certain format. The only requirement is to implement the method \ct{run:with:in:}.

\paragraph{Object replacement}

The primitive \ct{become: anotherObject} is provided by the VM and it atomically swaps the references of the receiver and the argument. All variables in the entire system that used to point to the receiver now point to the argument and vice-versa. In addition, there is also \ct{becomeForward: anotherObject} which updates all variables in the entire system that used to point to the receiver so that they point to the argument, \ie it is only one way.

This feature enables us to replace a target object with a proxy so that all variables that are pointing to the target object are updated to point to the proxy.

\section{Ghost's Design and Implementation}
\seclabel{ghostImpl}

Ghost is open-source and developed under the MIT license\footnote{\url{http://www.opensource.org/licenses/mit-license.php}}. The  website of the project with its documentation is in: \url{http://rmod.lille.inria.fr/web/pier/software/Marea/GhostProxies}. The source code is available in the SqueakSource3 server: \url{http://ss3.gemstone.com/ss/Ghost.html}.

\subsection{Overview Through an Example}

\begin{figure}[ht]\centering
  \includegraphics[width=0.6\linewidth]{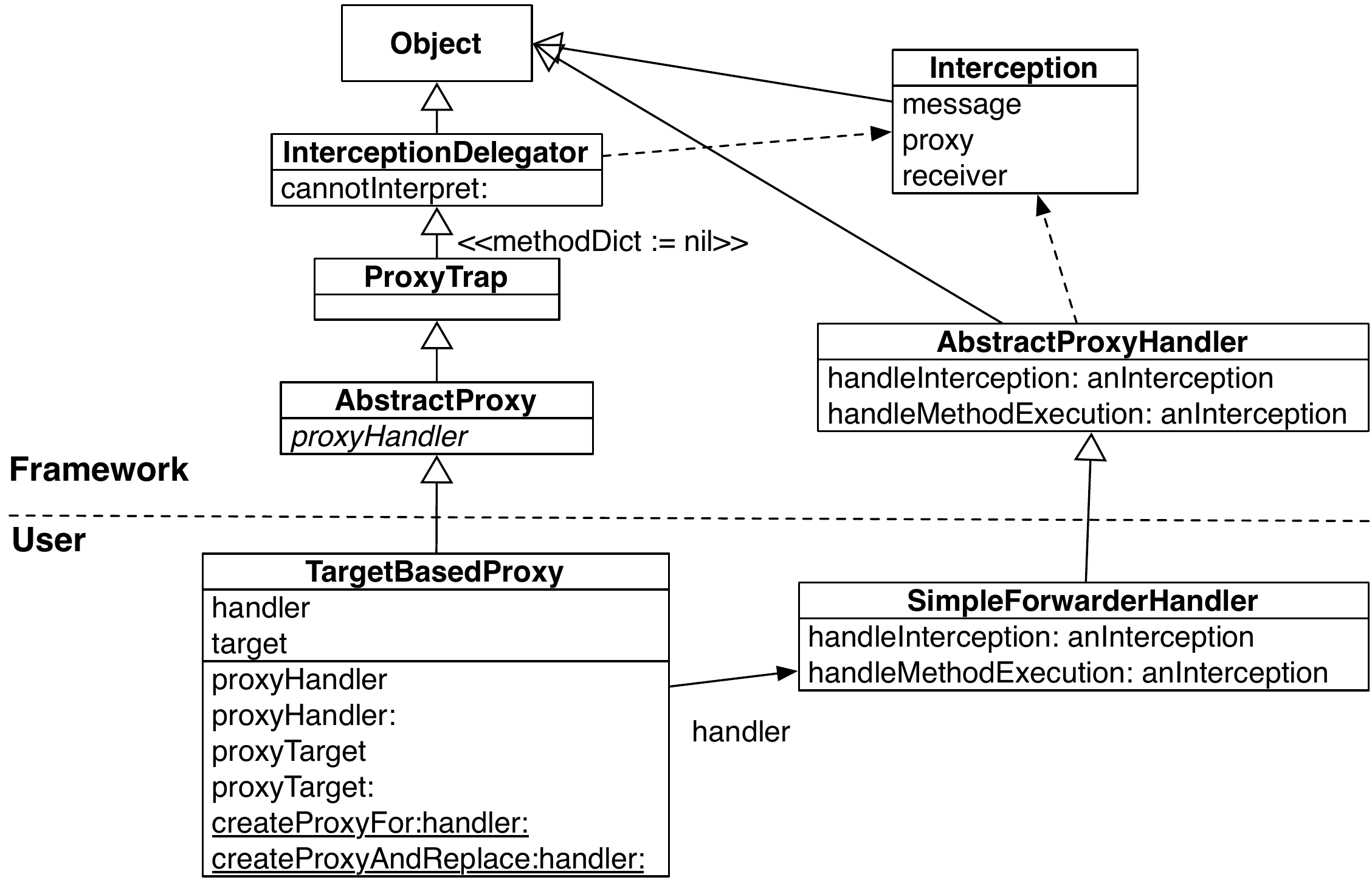}
  \caption{Part of the Ghost framework and an example of proxies for regular objects.}
  \label{fig:ghostStratifiedRegular}
\end{figure}

Ghost distinguishes between \emph{interceptors} and \emph{handlers}. Proxies solely play the role of interceptors while handlers define the treatment of the trapped message. Data related to a trapped message is reified as an object we call \emph{interception}. Figure~\ref{fig:ghostStratifiedRegular} shows Ghost's basic design which is explained in this section. The most important features of Ghost are: (1) to be able to intercept all messages but also to exclude a user-defined list, (2) to be uniform (to be able to proxify any objects, even sensitive ones like classes or method), and (3) to be stratified (\ie clear separation between proxies and handlers) in a \emph{controlled manner}.


Ghost's implementation uses the  following reflective facilities: classes with no method dictionary, objects as methods and object replacement.
The basic kernel is based on the hierarchies of \ct{AbstractProxy} (whose role is to intercept messages) and \ct{AbstractProxyHandler} (whose role is to handle intercepted messages) together with the communication from the former to the latter through \ct{Interception} instances.

The handlers' responsibility is to manage message interceptions trapped by proxies. What the handler does with the interception, depends on what the user wants. To illustrate the implementation, we use a \ct{SimpleForwarderHandler} which just forwards the intercepted message to a target object. In this example, each proxy instance uses a particular handler instance which is accessed by the proxy via an instance variable. Another user of the framework may want to use the same handler instance for all proxies. Consequently, different proxies can use the same or different handlers. How proxies are mapped to handlers depends on the user needs and it is controlled by the method \ct{proxyHandler} as explained later.

The information passed from a proxy to a handler is reified as an instance of the class \ct{Interception}. It includes the message which reifies the selector and its arguments, the proxy and the receiver (as we see later sometimes the receiver is not the proxy but a different object).

In real-world scenarios (see \secref{caseStudies}), a proxy often needs to hold some specific information, for example, a target object, an address in secondary memory, a filename, an identifier or any important information. Thus, the proxy should provide at least an accessor to allow the handler to retrieve this information. The need for these application-specific messages understood by the proxy leads us to \emph{controlled stratification}. In traditional proxy implementations the minimal object already understands some messages and therefore the developer cannot choose not to understand them. With our \emph{controlled stratification}, the \emph{developer controls} and decides the set (usually small) of messages understood by the proxy. By carefully choosing selectors for these messages (usually we use a specific prefix), one avoids collisions with applications messages and enhances proxy's transparency.

A typical reason for controlled stratification is saving memory by sharing a unique handler among all proxies. In the context of the simple forwarder example, we need to map each proxy to a target object that will eventually perform the messages trapped by the proxy. If one goes for full stratification, the proxy will intercept all messages and send them to the handler. But, the handler should hold a reference to the target. Then, for every target object we would have two placeholders: a proxy and a handler. If we use the same object for both responsibilities, then there is no clear division between proxies and handlers. Therefore, in this example, to have a smaller memory footprint, we introduced an instance variable in class \ct{TargetBasedProxy} that stores the target object. A singleton handler, shared among all proxies, asks each proxy for its target before forwarding  the intercepted message. To let the handler access the target object of a given proxy, \ct{TargetBasedProxy} class implements the method \ct{proxyTarget}. \secref{extendingProxies} explains that Ghost gives the user the flexibility to decide himself where and what data to store.

\noury{We need a discussion subsection about:\\
-Tension between transparency and genericity\\
-Why not:\\
--subclassing Interception and make the proxy use the subclass with IVs that are filled with the proxy state. Then we need to adapt proxy class (=> we lose genericity/reuse) to instantiate the appropriate interception class and fill it.\\
--proxy sends to handler a message which argument is an array with state. Still we need to adapt proxy class (=> we lose genericity/reuse) to build the array with the appropriate state}
\mmp{indeed, would be nice. I let it for the next version of the paper}

\subsection{Proxies for Regular Objects}

This section shows Ghost's implementation for regular objects. Subclasses of \ct{AbstractProxy} (such as \ct{TargetBasedProxy}) provide proxies for regular objects, \ie objects that do not need any special management. Their responsibility is to intercept messages.

\paragraph{Proxy creation} The following code shows how to create a proxy for a point (3,4). Since the handler is a simple forwarder, the messages are forwarded to the proxy's target.

\begin{code}{}
testSimpleForwarder
       | proxy |
       proxy := TargetBasedProxy createProxyFor: (Point x: 3 y: 4) handler: SimpleForwarderHandler new.
       self assert: proxy x equals: 3.
       self assert: proxy y equals: 4.
\end{code}

The class method \ct{createProxyFor:handler:} creates a new instance of \ct{TargetBasedProxy} and sets the handler (the user specifies which handler to use just by passing it as a parameter) and the target object.

\paragraph{Message Trapping in Action}
\ct{ProxyTrap} is a special class whose method dictionary is nilled out once created. When we send a message to an instance of \ct{TargetBasedProxy} if the message is not implemented in that class, the method lookup continues in the hierarchy until \ct{ProxyTrap}, whose method dictionary is \ct{nil}. For all those messages (the ones not implemented in \ct{TargetBasedProxy} and \ct{AbstractProxy}), the VM will eventually send the message \ct{cannotInterpret: aMessage}. Note that there are a few messages that are not executed but inlined by the compiler and the virtual machine (See \secref{specialMessages}). From now onwards, we consider that when we use \ct{cannotInterpret:}, we intercept \emph{all} messages except a specific list that we do not want to intercept. This is to distinguish it from \ct{doesNotUnderstand:} where one can only intercept the messages not understood.

Coming back to the \ct{cannotInterpret:}, remember that such message is sent to the receiver (in this case the \ct{aProxy} instance) but the method lookup starts in the superclass of the class whose  method dictionary is \ct{nil} which, in this case, is \ct{InterceptionDelegator} (see Figure~\ref{fig:ghostStratifiedRegular}). Because of this, \ct{InterceptionDelegator} implements the mentioned method:

\begin{code}{}
InterceptionDelegator >> cannotInterpret: aMessage
       | interception |
       interception := Interception for: aMessage proxy: self.
       ^ self proxyHandler handleInterception: interception.
\end{code}

An \ct{Interception} instance is created and passed to the handler. In this case, for the interception we only need the proxy and the message. The \ct{receiver} is unused here. \ct{InterceptionDelegator} sends \ct{proxyHandler} to get the handler. Therefore, \ct{proxyHandler} is an abstract method which must be implemented by concrete proxy classes, \eg \ct{TargetBasedProxy} and it must answer the handler to use.

Handler classes are user-defined and the example of the simple forwarder handler logs and forwards the received message to a target object as shown below.

\begin{code}{}
SimpleForwarderHandler >> handleInterception: anInterception
       | answer |
       self log: 'Message ', anInterception message , ' intercepted'.
       answer := anInterception message sendTo: anInterception proxy proxyTarget.
       self log: 'The message was forwarded to target'.
       ^ answer
\end{code}

Direct subclasses from \ct{AbstractProxy} \eg \ct{TargetBasedProxy} are only used for regular objects. We see in the following sections how Ghost handles objects that do require special management such as  methods (see \secref{proxyMethods}) or classes (see \secref{proxyClasses}).

\subsection{Extending and Adapting Proxies and Handlers}
\seclabel{extendingProxies}

To adapt the framework, users have to create their own subclass of \ct{AbstractProxyHandler} and implement the method \ct{handleInterception:}. They also need to subclass \ct{AbstractProxy} and define which handler to use by implementing the method \ct{proxyHandler}. It is up to the developer to store the handler in the proxy or to share a singleton handler instance among all proxies, or any other option. Other customizations are possible depending on the application's needs:

\begin{itemize}
\item  Which state to store in the proxy. For example, rather than a simple target object, proxies for remote objects may require an IP, a port and an ID identifying the remote object. A database or object graph swapper may need to store a secondary memory address or ID.
\item Which messages are implemented in the proxy and directly answered instead of being intercepted. The most common usage is implementing methods for accessing instance variables so that the handler can invoke them while managing an interception. Next section presents different examples.
\end{itemize}

\subsection{Intercepting Everything or Not?}

One would imagine that the best proxy solution is one that intercepts \emph{all} messages. However, this is not what the user of a proxy library needs most of the times. Usually, developers need to send messages to a proxy and get an answer instead of being intercepted.  Here are a few examples:

\begin{itemize}
\item Storing proxies in hashed collections means that proxies need to answer their hash.
\item With remote objects, it is likely that the system will need to ask a proxy its target in the remote system or information about it \eg URI or ID.
\item Serializing proxies to a file or network means that the serializer will ask its class and its instance variables to be serialized as well.
\item Debugging, inspecting and printing proxies only makes sense if a proxy answers its own information rather than intercepting the message.
\end{itemize}

The question ``Intercepting Everything or Not?''  is really a difficult one. On the one hand, to use a proxy as a placeholder, it is useful that it understands some basic messages such as \ct{identityHash}, \ct{inspect}, \ct{class}, etc. Not only the user of the proxy framework usually needs to send messages to a proxy, but also the proxy toolbox itself \eg \ct{proxyHandler}. On the other hand, it is a problem since those messages are not intercepted anymore.

To support these requirements, Ghost provides a flexible design so that proxies can understand and answer specific messages. The way to achieve this is simply by implementing methods in proxy classes. All methods implemented below \ct{ProxyTrap} in the hierarchy are not intercepted. With our solution, we have the best scenario: \emph{the user} controls stratification and decides \emph{what} to exclude in the proxies interception and intercept \emph{all} the rest. With solutions like \ct{doesNotUnderstand:}, one can also implement methods in proxy classes to avoid being intercepted but proxies are forced by the system to understand (and hence do \emph{not} intercept) even more messages like those methods that every object understands (e.g. \ct{identityHash}, \ct{initialize}, \ct{isNil}, etc.). Such messages are not defined by the user but by the system.

\subsection{Messages to be Answered by the Handler}

Apart from the possibility of adding methods to the proxy and avoiding interception of messages, Ghost supports special messages to which the \emph{handler} must answer itself instead of managing them as regular interceptions. A handler keeps a dictionary that maps selectors of messages intercepted by the proxy to selectors of messages to be performed by the handler. This user-defined list of selectors is used with different objectives such as debugging purposes, \ie those messages that are sent by the debugger to the proxy are answered by the handler and they are not managed as a regular interception. This eases the debugging in presence of proxies. The handler's dictionary of special messages for debugging can be defined as follows:

\begin{code}{}
SimpleForwarderHandler >> debuggingMessagesToHandle
      | dict |
      dict := Dictionary new.
      dict at: #basicInspect put:#handleBasicInspect:.
      dict at: #inspect put:#handleInspect:.
      dict at: #inspectorClass put:#handleInspectorClass:.
      dict at: #printStringLimitedTo: put: #handlePrintStringLimitedTo:.
      dict at: #printString put: #handlePrintString:.
      ^ dict
\end{code}

\mmp{maybe I should show some of these messages in the UML?}

The dictionary keys are selectors of messages received by the proxy and the values are selectors of messages that the handler must send to itself. For example, if the proxy receives the message \ct{printString}, then the handler sends itself the message \ct{handlePrintString:} and answers that.  All the messages to be sent to the handler (\ie the dictionary values) take as parameter an instance of \ct{Interception} which contains the message, the proxy and the receiver. Therefore, such messages have access to all the required information.

These special messages are pluggable \ie they are easily enabled and disabled. Moreover, they are not coupled with debugging so they can be used every time a user wants certain messages to be implemented and answered directly by the handler rather than performing the default action for an interception. As we explain in the next section, this feature is used, \eg to intercept method's execution.

This feature is similar to the ability of define methods in the proxy so that they are understood instead of intercepted. Nevertheless, there are some differences which help the user to decide which of the two ways to use in each situation:

\begin{itemize}
\item The mechanism of the handler is pluggable, while defining methods in a proxy is not.
\item Some methods like those accessing instance variables of the proxy (such as the target object) \emph{have} to be in the proxy.  Another example is primitive methods. For example, if we want the proxy to understand \ct{proxyInstVarAt:} so that it can be used for serialization purposes, we have to define such method in the proxy because it calls a primitive. It is impossible to define that method in the handler without changing the primitive.
\item Handlers can be shared among several proxy instances and even different types of proxies. Therefore, we cannot put specific behavior to a shared handler that applies only to a specific type of proxy.
\end{itemize}

\section{Proxies for Classes and Methods}
\seclabel{proxiesClassesAndMethods}

Ghost supports proxies for regular objects as well as for classes, methods and any other class that requires special management such as execution context, block closures or processes. In this section, we show how to correctly proxify methods and classes.

\subsection{Proxies for Methods}
\seclabel{proxyMethods}

In some dynamic languages, methods are first-class objects. This means that it is necessary to handle two typical and different scenarios when we want to proxify methods: (1) handling message sending to a proxified method object and (2) handling execution (from the VM) of a proxified method.

\begin{itemize}

\item \emph{Sending a message to a method object.} In Pharo, for example, when a developer searches for senders of a certain method, the system has to check in the literals of the compiled method if it is sending such message. To do this, the system searches all the literals of the compiled methods of all classes. This means it sends messages (\ct{sendsSelector:} in this case) to the objects that are in the method dictionary. When creating a proxy for a method, we need to intercept such messages.

\item \emph{Method execution.}  This is when the compiled method is executed by the virtual machine. Suppose we want to create a proxy for the method \ct{username} of \ct{User} class. We need to intercept the method execution, for example, when doing \ct{User new username}. Note that this scenario \emph{only} exists if a method is replaced by a proxy.

\end{itemize}

\paragraph{Proxy creation}
To clarify, imagine the following test:

\begin{code}{}
testSimpleProxyForMethods
       | mProxy kurt method |
       kurt := User named: 'Kurt'.
       method := User compiledMethodAt: #username.
       mProxy := TargetBasedProxy createProxyAndReplace: method handler: SimpleForwarderHandler new.
       self assert: mProxy getSource equals: 'username ^ name'.
       self assert: kurt username equals: 'Kurt'.
\end{code}

The test creates an instance of \ct{User} class and a proxy for method \ct{username}. Using the method \ct{createProxyAndReplace:handler:}, we create the proxy and we \emph{replace} the original object (the method \ct{username} in this case) with it. By replacing an object, we mean that all the pointers to the existing method then point to the proxy. Apart from \ct{createProxyAndReplace:handler:}, the proxy also understands \ct{createProxyFor:handler:} which creates the proxy but does replace objects, \ie object replacement is optional in Ghost.

Finally, we test both types of messages:  sending a message to the proxy (in this case \ct{mProxy getSource}) and executing a proxified method (\ct{kurt username} in our example).

\paragraph{Handling both cases}
Ghost solves both scenarios. In the first one, \ie ~\ct{mProxy getSource}, Ghost has nothing special to do. It is just a message sent to a proxy and it behaves exactly the same way we have explained so far. In the second one, illustrated by \ct{kurt username}, a proxified method is \emph{executed}. In this case, Ghost uses the reflective capability ``objects as methods'', \ie when the VM looks for the method \ct{username}, it notices that, in the method dictionary, there is not a \ct{CompiledMethod} instance but instead an instance of another class. Consequently, it sends the message \ct{run:with:in} to such object. Since such object is a proxy in this case, the message \ct{run:with:in:} is intercepted and delegated to the handler just like any other message.

As already explained, the handler can have a list of messages that require special management rather than performing the default action. With that feature, we map \ct{run:with:in} to \ct{handleMethodExecution:}, meaning that if the handler receives an interception with the selector \ct{run:with:in} it sends to itself \ct{handleMethodExecution:} and answers that. Subclasses of \ct{AbstractProxyHandler} that want to handle interceptions of methods' execution must implement \ct{handleMethodExecution:} to fit their needs, for example:

\begin{code}{}
SimpleForwarderHandler >> handleMethodExecution: anInterception
     | targetMethod receiver arguments |
     targetMethod := anInterception proxy proxyTarget.
      "Remember the message was run: aSelector with: arguments in: aReceiver"
     receiver := anInterception message arguments third.
     arguments := anInterception message arguments second.
     ^ targetMethod valueWithReceiver: receiver arguments: arguments
\end{code}

That method just gets the required data from the interception and executes the method with the correct receiver and arguments by using the method \ct{valueWithReceiver:arguments:}.

Notice that the Pharo VM does not impose any shape to methods. Therefore, as we showed in the previous example, we can use the same proxy class (\ct{TargetBasedProxy}) that we use for regular objects.

\paragraph{Alternatives} Another approach to manage interceptions of methods' execution is to implement \ct{run:with:in:} in the proxy itself. In such situation, we can get the data from the parameters, create an \ct{Interception} instance and pass it to the handler. However, we believe proxies should understand as little as possible from the proxy toolbox machinery and leave such responsibilities for their handlers.

\subsection{Proxies for Classes}
\seclabel{proxyClasses}

Pharo represents classes as first-class objects and they play an important role in the runtime infrastructure. It is because of this that Ghost has to take them into account. Developers often need to be able to replace an existing class with a proxy. Instances hold a reference to their class and the VM uses this reference for the method lookup. Therefore, object replacement must not only update the references from other objects, but also the class references from instances.

 \begin{figure}[ht]\centering
     \includegraphics[width=0.8\linewidth]{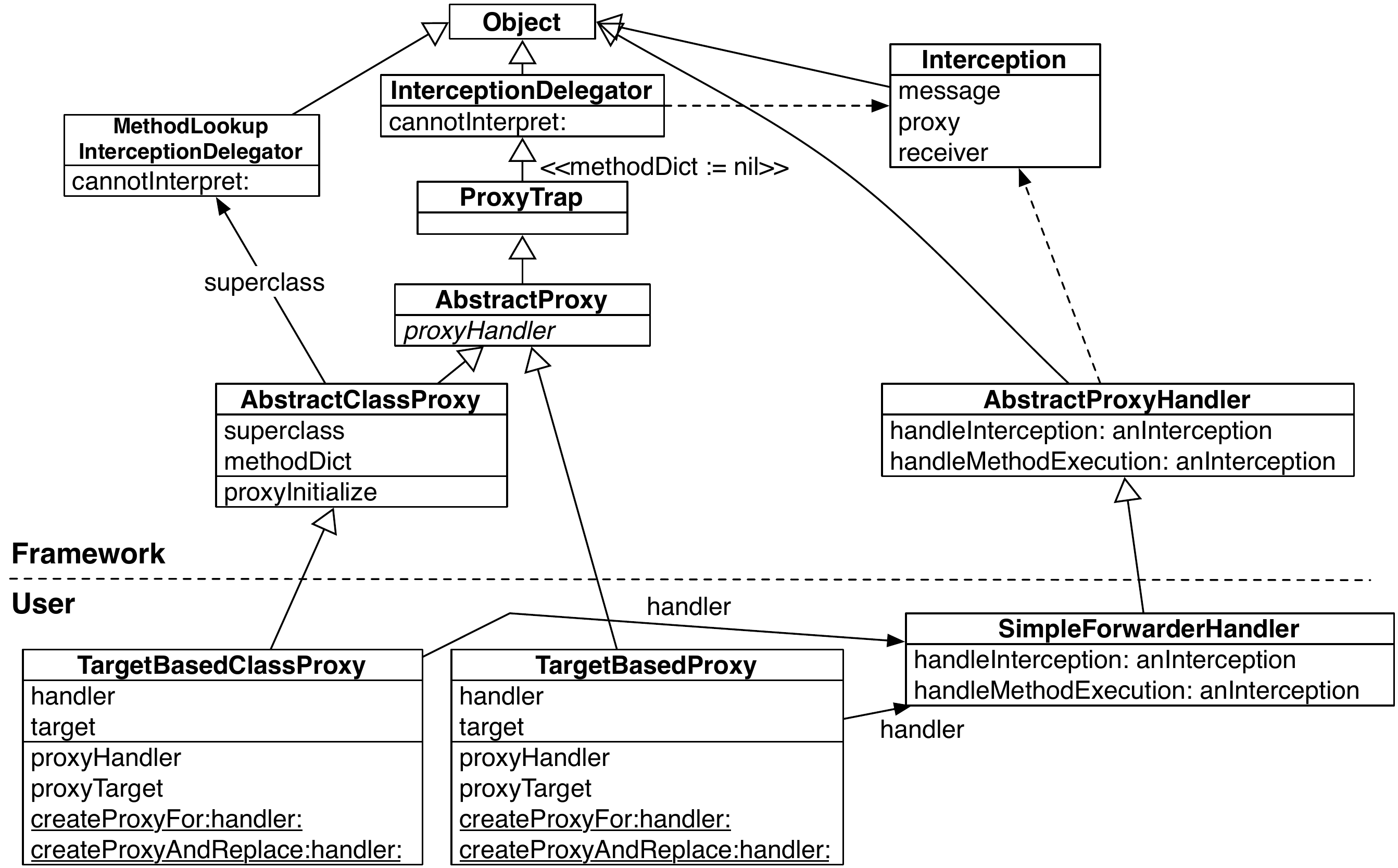}
     \caption{Part of the Ghost framework and an example of proxies for classes.}
     \label{fig:ghostStratified2}
\end{figure}

\ct{AbstractClassProxy} provides the basis for class proxies (See Figure~\ref{fig:ghostStratified2}). \ct{AbstractClassProxy} is necessary because the VM imposes specific constraints on the memory layout of objects representing classes. The VM expects a class object to have the two instance variables \ct{superclass} and \ct{methodDict} in this specific order starting at index 1. We do not want to define \ct{TargetBasedClassProxy} as a subclass of \ct{TargetBasedProxy} because the two instance variables \ct{target} and \ct{handler} would get index 1 and 2, not respecting the imposed order.  However, not being able to subclass is not a real problem in this case because there are only a few methods in common so we are not duplicating much code because of that.

\paragraph{Requirements}
\ct{AbstractClassProxy} has to be able to intercept the following kinds of messages:

\begin{itemize}
\item Messages that are sent directly to the class as a regular object.
\item Messages that are sent to an instance of the proxified class, \ie an object whose class reference is pointing to the proxy (which happens as a consequence of the object replacement between the class and the proxy). This kind of message is only necessary when an object replacement takes place.
\end{itemize}

\paragraph{Proxy creation} To explain class proxies, consider the following test:

\begin{code}{}
testSimpleProxyForClasses
       | cProxy kurt |
       kurt := User named: 'Kurt'.
       cProxy := TargetBasedClassProxy createProxyAndReplace: User handler: SimpleForwarderHandler new.
       self assert: User name equals: #User.
       self assert: kurt username equals: 'Kurt'.
\end{code}

 This test creates an instance of \ct{User} and then, with the message \ct{createProxyAndReplace:handler:}, it creates a proxy that replaces the \ct{User} class. Finally, it tests that we can intercept both kind of situations: messages sent to the proxy (in this case \ct{User name}) and messages sent to instances of the proxified class (\ct{kurt username} in this case).

\sd{why createProxyAndReplace:handler: is not part of the framework?}
\mmp{because that depends on the user, not all proxy classes need a handler to be passed as argument. For example, in Marea I could implement proxyHandler as returning MareaHandler uniqueInstance... so.. I don't need the handler at all there...}

\paragraph{Handling two cases}
The first message, \ct{User name}, has nothing special and it is handled like any other message. The VM will end up sending \ct{cannotInterpret:} to the receiver and starting the method lookup in the class which method dictionary was \ct{nil}, \ie ~\ct{InterceptionDelegator}. The second message is more complicated and needs certain explanation.

The method \ct{createProxyAndReplace:handler:} is equal to the one of \ct{AbstractProxy} but something different happens. After creating a new instance, the method \ct{createProxyAndReplace:handler:} sends the message \ct{proxyInitialize} to it. In \ct{TargetBasedClassProxy} we do not only set a handler and a target (as in the case of \ct{TargetBasedProxy}), but also the minimal information required by the VM so that an \emph{instance} of \ct{TargetBasedClassProxy} can act as a \emph{class}. Thus, we set its method dictionary to \ct{nil} and its superclass to \ct{MethodLookupInterceptionDelegator}.

Coming back to the example, when we evaluate \ct{kurt username}, the class reference of \ct{kurt} is pointing to the created \ct{TargetBasedClassProxy} instance (as a result of object replacement). This proxy object acts as a class and it has its method dictionary instance variable to \ct{nil}. Hence, the VM sends the message \ct{cannotInterpret:} to the receiver (\ct{kurt} in this case) but starting the method lookup in the superclass of the class with nilled method dictionary which is \ct{MethodLookupInterceptionDelegator}.
\noury{You need a drawing here to show the object graph after the become and control flow as you did in figure 6}
\mmp{noury, you comment it very old and I dont know now which is the kind of figure you are talking about :( }
A simplified definition (later in this section we see the real implementation) of the \ct{cannotInterpret:} of class \ct{MethodLookupInterceptionDelegator} is the following:

\begin{code}{}
MethodLookupInterceptionDelegator >> cannotInterpret: aMessage
      | proxy |
      proxy := aMessage lookupClass.
      interception := Interception for: aMessage proxy: proxy receiver: self.
      ^ proxy proxyHandler handleInterceptionToInstance: interception.
 \end{code}

It is important to notice the difference in this method in comparison with the implementation of \ct{InterceptionDelegator}. In both situations, \ct{User name} and \ct{kurt username}, Ghost always needs to get the proxy to perform the desired action.

\begin{description}
\item \ct{User name} case.
The method \ct{cannotInterpret:} is called on \ct{InterceptionDelegator} and the receiver, \ie what \ct{self} is pointing to, is the proxy itself.

\item  \ct{kurt username} case.
The method \ct{cannotInterpret:} is called on \ct{MethodLookupInterceptionDelegator} and \ct{self} points to \ct{kurt} and not to the proxy. The proxy is the looked up class, \ie the receiver's class,  which we can get from the \ct{Message} instance. Then we send the message \ct{handleInterceptionToInstance:} to the handler. We use that message instead of \ct{handleInterception:} because the user may need to perform different actions. What the implementation of \ct{handleInterceptionToInstance:} does in \ct{SimpleForwarderHandler} is to execute the desired method with the receiver without sending a message to it\footnote{Pharo provides the primitive method  \ct{receiver:withArguments:executeMethod:} which directly evaluates a compiled method on a receiver with a specific list of arguments without actually sending a message to the receiver.} avoiding another interception and an infinite loop.

\end{description}

To conclude, with this implementation, we can successfully create proxies for classes, \ie to be able to intercept the two mentioned kind of messages and replace classes by proxies.

\paragraph{Discussion} We could have reused  \ct{cannotInterpret:} implementation of \ct{InterceptionDelegator} instead of using \ct{MethodLookupInterceptionDelegator} and set it also in the method \ct{proxyInitialize} of \ct{AbstractClassProxy}. That way,  \ct{InterceptionDelegator} is taking care of both types of messages. The solution has to check which kind of message it is and, depending on that, perform a specific action. We think the solution with \ct{MethodLookupInterceptionDelegator} is much cleaner.

Ghost's implementation uses \ct{AbstractProxyClass} not only because it is cleaner from the design point of view, but also because of the memory footprint.  Technically, we can \emph{also} use \ct{AbstractProxyClass} for regular objects and methods. However, this implies that, for every target to proxify, the size of the proxy is unnecessary bigger in memory footprint because of the additional instance variables needed by \ct{AbstractProxyClass}.

\paragraph{Problem with subclasses of proxified classes}
\noury{Add a drawing illustrating the problem}
When we proxify a class but not its instances and one of the instances receives a message, Ghost intercepts the method lookup and finally uses the \ct{cannotInterpret:} method from \ct{MethodLookupInterceptionDelegator}. In that method, the proxy can be obtained using \ct{aMessage lookupClass}, because the class of the receiver object is a proxy. However, this is not always possible. If we proxify a class but not its subclasses and a subclass' instance receives a message which does not match any method, the lookup eventually reaches the proxified class. Ghost intercepts the method lookup and executes the \ct{cannotinterpret:} method from \ct{MethodLookupInterceptionDelegator}. At this stage, we need to \emph{find} the trapping class, i.e., the first class in the hierarchy with a nilled method dictionary. In this scenario, \ct{message lookupClass} does not return a proxy but an actual class: a subclass of the proxified class. To solve this problem, Ghost does the following implementation:

\begin{code}{}
MethodLookupInterceptionDelegator >> cannotInterpret: aMessage
      | proxyOrClass proxy |
      proxyOrClass := aMessage lookupClass.
      proxy := proxyOrClass ghostFindClassWithNilMethodDictInHierarchy.
      interception := Interception for: aMessage proxy: proxy receiver: self.
      ^ proxy proxyHandler handleInterceptionToInstance: interception.
 \end{code}

The method \ct{ghostFindClassWithNilMethodDictInHierarchy} checks if the current class has a nilled method dictionary and, if it does not, it recurs to superclasses. This method is also implemented in \ct{AbstractClassProxy} just answering \ct{self}.

\section{Special Messages and Operations}
\seclabel{specialMessages}

Being unable to intercept messages is a problem because it means they will be directly executed by the proxy instead. This can lead to different execution paths in the code, errors or even make the VM to crash \cite{Aust11a}.

One common problem when trying to intercept all messages is the existing optimizations for certain methods. In Pharo, as well as in other languages, there are two kinds of optimizations that affect proxies: (1) inlined methods and (2) special bytecodes.

\subsection{Inlined Methods}

These are optimizations done by the compiler. For example, messages like \ct{ifTrue:}, \ct{ifNil:}, \ct{and:}, \ct{to:do:}, etc. are detected by the compiler and are not encoded with the regular bytecode of message send. Instead, such methods are directly encoded using different bytecodes such as jumps. As a result, these methods are never executed and cannot be intercepted by proxies. The second kind of optimization is between the compiler and the virtual machine.

Ideally, we would like to handle inlined messages the same way than regular ones. The easiest yet naive way is to disable the inlining. However, disabling all optimizations produces two important problems. First, the system gets significantly slower.
Second, when optimizations are disabled, those methods are executed and there can be unexpected and random problems which are difficult to find. For instance, in Pharo, everything related to managing processes, threads, semaphore, etc., is implemented in Pharo itself without proper abstractions. The processes' scheduler can only switch processes between message sends. This means that there are some parts in the classes like \ct{Process}, \ct{ProcessorScheduler}, \ct{Semaphore}, etc., that have to be atomic, \ie they cannot be interrupted and switched to another process. If we disable the optimizations, such code is not atomic anymore. Other examples are the methods used to enumerate objects or to get the list of objects pointing to another one. While iterating objects, each send to \ct{whileTrue:} (or any other of the inlined methods) will create more objects like \ct{MethodContext} generating an infinite loop.

The messages that are inlined in Pharo 1.4 are stored in the class variable \ct{MacroSelectors} of the class \ct{MessageNode} and they are:
\begin{enumerate}
\item \ct{ifTrue:}, \ct{ifFalse:}, \ct{ifTrue:ifFalse:}, \ct{ifFalse:ifTrue:}, \ct{and:}, \ct{or:},  implemented in \ct{True} and \ct{False}.
\item  \ct{ifNil:}, \ct{ifNotNil:}, \ct{ifNil:ifNotNil:} and \ct{ifNotNil:ifNil:},  implemented in both classes: \ct{ProtoObject} and \ct{UndefinedObject}.
\item \ct{to:do:} and \ct{to:by:do:},  implemented in \ct{Number}.
\item \ct{whileFalse:}, \ct{whileTrue:}, \ct{whileFalse}, \ct{whileTrue} and \ct{repeat},  implemented in \ct{BlockClosure}.
\end{enumerate}

These messages involve special objects that can be split into two categories.
In the first category, we have \ct{true}, \ct{false}, \ct{nil}, and numbers which are related to the messages of items 1 to 3.
These objects are so low-level that they cannot be replaced by proxies.
Indeed, sending the \ct{become:} message to one of them result into VM hang or crash.
To our knowledge there is no use case where these objects should be proxified.
Hence, inlining their messages is not an issue.

Block closures form the second category of objects which are involved in inlined messages.
Actually, the inlining of the messages \ct{whileFalse:}, \ct{whileTrue:}, \ct{whileFalse}, \ct{whileTrue} and \ct{repeat} is performed only when the receiver is a closure.
In situations where the receiver is the result of a message or a variable like in the code below, the message is not inlined.
The following code illustrates these two cases:

\begin{code}{}
	|welcomeBlock|
	[Transcript cr; show: 'Hello'] repeat.  "Message inlined"
	welcomeBlock := [Transcript cr; show: 'Welcome'].
	welcomeBlock repeat.  "No inlining"
\end{code}

Since a proxy is not recognized as closure by the compiler, the message is not inlined. Therefore, messages sent  to a proxy for a block closure are intercepted.

 \subsection{Special Associated Bytecodes and VM optimization}

The second type of optimization correspond to a special list of selectors\footnote{In Pharo 1.4, we can get the list of those selectors by executing \ct{Smalltalk specialSelectors}. Those are: \ct{+}, \ct{-}, \ct{<}, \ct{>}, \ct{<=}, \ct{>=}, \ct{=}, \texttt{\small \textasciitilde=}, \ct{*}, \ct{/},  \ct{\textbackslash\textbackslash}, \ct{@}, \ct{bitShift:}, \ct{//}, \ct{bitAnd:}, \ct{bitOr:}, \ct{at:}, \ct{at:put:}, \ct{size}, \ct{next}, \ct{nextPut:}, \ct{atEnd}, \ct{==}, \ct{class}, \ct{blockCopy:}, \ct{value}, \ct{value:}, \ct{do:}, \ct{new}, \ct{new:}, \ct{x} and \ct{y}} that the compiler does not encode with the regular bytecode of message send. Instead, these selectors are associated with special bytecodes that the VM directly interprets. For these selectors, there are three groups:

\begin{itemize}

\item Methods that may be executed depending on the receiver or argument type. For example, the execution \ct{1 + 2} never sends the message \ct{+} to 1 because both are 32 bit integers but \ct{1 + 'aString'} will do. Analyzing the implementation of the bytecode associated with each of those selectors shows us that all of them check the type of the receiver and arguments: the method is only executed when there is a type mismatch.
For example, all arithmetic operations and bit manipulation expect small integers or floats, boolean operations expect booleans, \ct{size} expects strings or arrays, etc.  Whenever the receiver or arguments do not satisfy the conditions, the bytecode follows with the normal method execution, \ie, the message is sent. Since proxies never satisfy the conditions, then the messages are sent by the VM and trapped like normal messages.

\item  Methods that are always sent such as \ct{new}, \ct{next}, \ct{nextPut:}, \ct{do:}, etc. Here the only optimization done by the VM is just a quick and internal set of the selector to execute and the argument count. These methods are not a problem for proxies since they are always executed.

\item Methods which are never executed but directly answered by the VM internal execution.
In Pharo 1.4, there is only one single method of this type\footnote{In earlier versions of Pharo \ct{class} was also of this type, but the special bytecode associated it was removed in Pharo 1.4.}:
\ct{==}. It answers whether the receiver and the argument are the same object.
\end{itemize}

The conclusion is that only \ct{==} is not intercepted by proxies. Nevertheless, even if it were possible, object identity should never be handled as a regular message as demonstrated by Mark Miller \footnote{http://erights.org/elib/equality/grant-matcher/index.html}. However, object replacement conflicts with object identity. For example, given the following code:

\begin{code}{}
 (anObject  == anotherObject)
         ifTrue: [ self doSomething]
         ifFalse: [self doSomethingDifferent]
\end{code}

Imagine that \ct{anObject} is replaced by a proxy, \ie all objects in the system which were referring to the target (\ct{anObject}), will now refer to the proxy.  Since all references have been updated, \ct{==} continues to answer correctly. For instance, if \ct{anotherObject} was the same object as \ct{anObject}, \ct{==} answers true since both are referencing the proxy now. If they were not the same object, \ct{==} answers false. Hence, checking identity is not a problem when there is object replacement. However, there is more to object identity than testing for equality. For example, object identity is changed with object replacement and this can affect security, \eg trusted objects can become untrusted or vice-versa. This is one of the reasons why object replacement is optional in Ghost.

\section{Criteria Evaluation}
\seclabel{criteriaGhost}

In this section, we evaluate Ghost using the criteria defined in \secref{proxySurvey}.

\paragraph{Stratification}  This solution is stratified. On the one hand, there is a clear separation between proxies and handlers. On the other hand, interception facilities are separated from application functionality. Indeed, the application can even send the \ct{cannotInterpret:} message to the proxy and it will be intercepted like any other message. Thus, the proxy API does not pollute the application's namespace. Moreover, stratification is controlled: users can still select which messages they do not want to be intercepted.

\paragraph{Interception levels}  It can intercept all messages while also providing a way to exclude user defined messages.

\paragraph{Object replacement} Such feature is important since it allows one to seamlessly substitute an object with a proxy and this is provided by Ghost thanks to the \ct{become:} primitive of Pharo.  However, object replacement is optional meaning the user can decide whether to replace objects with proxies or not.

\mmp{we still have the problem of reference-leak as explain in the criteria evaluation of the common proxy implementation}

\paragraph{Uniformity}  This implementation is quite uniform since proxies can be used for regular objects as well as for classes and methods. Ghost does not yet provide out-of-the-box proxies for processes, contexts and blocks. However, it does provide the infrastructure and flexibility to create special proxies for them when needed.

All proxies provide the same API and can be used polymorphically. Nevertheless,  there is still non-uniformity regarding some other special classes and objects. Most of them are those that are present in what is called the \emph{special objects array}\footnote{Check the method \ct{recreateSpecialObjectsArray} in Pharo for more details.} which contains the list of special objects that are known by the VM. Examples are the objects \ct{nil}, \ct{true}, \ct{false}, etc. It is not possible to do a correct object replacement of those objects with proxies. The same happens with immediate objects, \ie objects that do not have object header and are directly encoded in the memory address such as \ct{SmallInteger}.

The special objects array contains not only regular objects, but also classes. Those classes are known and used by the VM so it may impose certain shape, format or responsibilities in their instances. For example, one of those classes is \ct{Process} and it is not possible to correctly replace a \ct{Process} instance by a proxy. 
These limitations occur only when object replacement is desired. Otherwise, there is no problem and proxies can be created for those objects.
Since classes and methods play an important role in the runtime infrastructure of Pharo, creating proxies for them useful in several scenarios (as we see in the next section) and that is why Ghost provides special management for them. From our point of view, the mentioned limitations only exist in the presence of unusual needs. Nevertheless, if the user also needs special management for certain objects like \ct{Process} instances, then he can create a particular proxy that respects the imposed shape.

\paragraph{Transparency}  Ghost proxies are transparent even with the special messages inlined by the compiler and the VM. The only exception is object identity (message \ct{==}).

In \secref{criteria} we mention the problem of ``target leaking'' in which a proxy returns a reference to the target object as an answer to a message. Ghost does not impose any particular state in the proxies and so they may not even have a target object. However, in the example of the \ct{SimpleForwarderHandler}, which indeed has a target, Ghost did solve this problem. When the handler processes an interception it forwards the message to the target object. Then the handler checks whether the answer from the target to that message was the target itself. If it was, then the handler answers the proxy, otherwise the original answer.

\paragraph{Efficiency}

Since Ghost interception is also based on two lookups (one for the original message and one for the \ct{cannotInterpret:}) and the mechanism is similar, it has the same performance than the traditional proxies.

However, Ghost provides an efficient memory usage with the following optimizations:

\begin{itemize}

\item \ct{TargetBasedProxy} and \ct{TargetBasedClassProxy} are defined as compact classes.
This means that in a 32 bits system, their instances' object header is only 4 bytes long instead of 8 bytes. For those instances whose body part is more than 255 bytes and whose class is compact, their header is 8 bytes instead of 12.
The first word in the header of regular objects contains flags for the garbage collector, the header type, format, hash, etc.
The second word is used to store a reference to the class.
In compact classes, the reference to the class is encoded in 5 bits in the first word of the header.
These 5 bits represent the index of a class in the compact classes array set by the language\footnote{See methods \ct{SmalltalkImage>>compactClassesArray} and \ct{SmalltalkImage>>recreateSpecialObjectsArray}.} and accessible from the VM.
With these 5 bits, there are 32 possible compact classes. This means that, from the language side, the developer can define up to 32 classes as compact. Declaring the proxy classes as compact, allows proxies to have smaller header and, consequently, smaller memory footprint.

\item Proxies only keep the minimal state they need. \ct{AbstractProxy} defines no structure and its subclasses may introduce instance variables needed by applications.

\item In the methods for creating proxies presented so far (\ct{createProxyFor:handler:} and \ct{createProxyAndReplace:handler:}), the last parameter is a handler. This is because, in our example, each proxy holds a reference to the handler. However, this is only necessary when the user needs one handler instance per target object which is not often the case. The handler is sometimes stateless and can be shared and referenced through a class variable or a global one. In that scenario, \ct{proxyHandler} must be implemented to answer a singleton. Therefore, we can avoid the memory cost of a \ct{handler} instance and its reference from the proxy. If we consider that the handler has no instance variable, then it is 4 bytes saved for the instance variable in the proxy and 8 bytes for the handler instance. That gives a total of 12 bytes saved per proxy in a 32 bits.

\end{itemize}

\paragraph{Ease of debugging}  Because Ghost provides a way to have messages answered directly by the handler, we can make debugging with proxies very easy. The handler can answer all methods related to debugging, inspecting, etc. In contrast with traditional proxy implementation based on \ct{doesNotUnderstand:}, this mechanism is pluggable, \ie it can be enabled or disabled at runtime by just changing a dictionary. There is no need to remove or add methods to the proxy.

\paragraph{Implementation complexity} Ghost is easy to implement in Pharo: it consists of 11 classes with an average of 6 methods per class and each method has an average of 4 lines of code. The total amount of lines of code is approximately 300. Ghost is covered by approximately 10 unit tests that cover all use cases.

\paragraph{Constraints} The solution is flexible since the objects to proxify can inherit from any class and are free to implement or not all the methods they want. There is no kind of restriction imposed by Ghost. In addition, the user can easily extend or change the purpose of the toolbox adapting it to his own needs by just subclassing a handler and a proxy.

\paragraph{Portability} Ghost is not portable to other Smalltalk dialects because it is based on a VM hook (the \ct{cannotInterpret:} message) present in the Pharo VM. In addition, it also needs object replacement (\ct{become:} primitive) and objects as methods (\ct{run:with:in:} primitive). The \ct{become:} primitive is present in all Smalltalk dialects because it is used by the language itself. The hook method \ct{run:with:in:} is not available in all dialects but we only need it if we need to intercept method execution. Furthermore, we rely on the primitive method \ct{receiver:withArguments:executeMethod:} to be able execute a method on a receiver object without actually sending a message to it. We only use this method when we proxify classes and this method is present in some Smalltalk dialects.

Without these reflective facilities, we cannot easily implement all the required features of a good proxy library. In the best scenario, we can do it but with substantial development effort such as modifying the VM or compiler or even creating them from scratch. Pharo provides all those features by default and no changes are required for Ghost.

\section{Discussion}
\seclabel{discussion}

Ghost applicability does not depend solely on the particular way in which Pharo handles the \ct{cannotInterpret:} mechanism. This is just the hook Ghost uses to intercept messages instead of the traditional \ct{doesNotUnderstand:} based proxies. The \ct{cannotInterpret:} primitive is needed by only a few features of those provided by Ghost, \ie stratification and the ability to intercept all messages.

The rest of the features, contributions and problems solved by the Ghost model can also be implemented with different hooks to intercept messages \eg with the \ct{doesNotUnderstand:}. The clear division between proxies and handlers, the special messages in the handler that it can answer itself, the ease of debugging, the ability to proxify methods and classes, to name a few, are independent of the hook used to intercept messages. This means that the Ghost model could be implemented in another Smalltalk dialect and get most of the features provided right now by the Pharo implementation. Therefore, Ghost proposal is still valuable and a  contribution to classical proxies.

The ability to proxify methods and classes is not mandatory in Ghost. The user can still use the framework for regular proxies and obtain all the mentioned advantages. However, if proxies for classes and methods is needed, then the implementation language must support a way to intercept method execution and a way to deal with objects whose classes are proxies.  Pharo provides us both of them.

\section{Case Studies}
\seclabel{caseStudies}

As a matter of showing possible uses of Ghost proxies, we present the following real cases:

\subsection{Marea}

\mmp{put paper of JOT if it is accepted!!!}

In OO software, some objects are only used in certain situations or conditions and remain not used for a long period of time.
We qualify such objects as  \emph{unused}.
These objects are reachable, and thus cannot be garbage collected. This is an issue because unused objects waste primary memory \cite{Kaeh86a}.

Operating systems have been supporting virtual memory since a long time \cite{Denn70a,Carr81a}. Virtual memory is transparent in the sense that it automatically swaps out unused memory organized in pages governed by some strategies such as the least-recently-used  (LRU) \cite{Chu72a}.
As virtual memory is transparent, it does not know the application's memory structure, nor does the application have any way to influence the virtual memory manager.

\subsubsection{Marea Overview}

Marea, is a virtual memory manager whose main goal is to offer the programmer a solution to handle application-level memory \cite{mart12a}.
Developers can instruct the system to release primary memory by swapping out \emph{unused objects} to secondary memory \cite{Mart11b,Mart11c}.
Marea is designed to: 1) save as much memory as possible \ie the memory used by its infrastructure is minimal compared to the amount of memory released by swapping out unused objects,
2) minimize the runtime overhead \ie  the swapping process is fast enough to avoid slowing down primary computations of applications, and
3) allow the programmer to control or influence the objects to swap.

The input for Marea are object graphs. These graphs are serialized and moved to secondary memory. Swapped out graphs are swapped in (read from secondary memory and materialized) as soon as one of their elements is needed. Graphs to swap can have \emph{any} shape and contain \emph{any} kind of object. This includes classes, methods, closures and even the execution stack which are all first-class objects in Pharo, Marea's implementation language.

When Marea swaps a graph, it correctly handles all the references from outside and inside the graph. When one of the swapped objects is needed, its graph is \emph{automatically} brought back into primary memory. To achieve this, Marea replaces original objects with proxies (object replacement). Whenever a proxy intercepts a message, it loads back the swapped graph from secondary memory.  This process is completely transparent for the developer and the application, \ie any interaction with a swapped graph has the same results as if it was never swapped.

Marea proxifies and serializes regular objects, methods and classes. This is a challenge since it requires special handling. For the serialization, Marea uses Fuel \cite{Dias12a}, a fast binary object graph serializer. For the proxies toolbox, Marea uses Ghost.

\subsubsection{Marea Proxies and Handlers}

Marea has its own subclasses of \ct{AbstractProxy} which do not store a target object but instead, a proxy ID (composed by a graph ID and a position) which is needed by the algorithms to swap out and in. When a proxy intercepts a message, it means that the swapped-out object graph is needed again. Because of this, for every interception \ct{MareaProxyHandler} (subclass  of \ct{AbstractProxyHandler}) reads the swapped-out object graph from secondary memory (the proxy has the needed information) loading it into primary memory and resending to it the original intercepted message. The following method illustrates this behavior:

\begin{code}{}
MareaProxyHandler >> handleInterception: anInterception
       | originalObject |
       originalObject := self loadFromSecondaryMemory: anInterception proxy.
       ^ anInterception message sendTo: originalObject.
\end{code}

\subsubsection{Requirements and Advantages of Using Ghost}

\paragraph{Proxifying methods and classes}

Typical unused objects are part of the applications' runtime. For example, classes and their methods are loaded on startup but most of them are useless regarding application functionalities. Consequently, applications usually occupy more memory than they actually need. Therefore, Marea needs Ghost to be able to proxify classes and methods.

\paragraph{Reducing memory occupy by proxies}

Since for Marea it is important that proxies have the minimum memory footprint possible, it takes advantage of some features provided by Ghost:
\begin{itemize}
\item Proxies are instances of \emph{compact classes}.
\item Since \ct{MareaProxyHandler} is stateless, it is shared among proxies.
\item Marea encodes the proxy instance variables \ct{position} and \ct{graphID} in one unique \ct{proxyID}. The \ct{proxyID} is a \ct{SmallInteger} which uses 15 bits for the \ct{graphID} and 16 bits for the \ct{position}\footnote{Marea also provides large proxies when the limits are exceeded.}.  Since \ct{SmallInteger} are immediate objects in Pharo, there is no need for an object header for the \ct{proxyID}.
\end{itemize}

\paragraph{Avoiding unnecessary swap-ins}

Having classes and methods as first-class objects offers solid reflective capabilities. However, some system queries access \emph{all} classes or methods in the system, that may cause the swap in of many of the swapped out graphs. These scenarios happen during application development. As Marea is intended to reduce memory for deployed applications, this is not usually a problem. Still, Marea provides a solution thanks to Ghost capabilities.

The solution requires to have certain messages handled by the proxy itself instead of forwarding it to the handler (that will swap in the graph). The proxy plays the role of a cache by keeping certain information of the proxified object. For example, Marea uses it for the case of class proxies. In Pharo, it is common that the system simply select classes by sending messages such as  \ct{isBehavior}, \ct{isClassSide}, \ct{isInstanceSide}, \ct{instanceSide} or \ct{isMeta}. Therefore, Marea defined such methods in \ct{ClassProxy} to answer appropriate results, \ie~\ct{isBehavior} and \ct{isInstanceSide} answers true, \ct{isClassSide} and \ct{isMeta} answers false and \ct{instanceSide} answers \ct{self}. This way, Marea avoids swapping in classes.

Marea also applied this solution to metaclasses and traits. But, it generalizes to any kind of object.  An improvement of the previous solution, is to make proxies cache some values from the proxified objects. For instance, a class proxy can cache the class name, a method proxy can cache the literals of the proxified method. There is a trade-off here between sizes of proxies and proxified objects.
It is often worth it to have larger proxies if they avoid swapping in large objects or objects that are roots of relatively large object subgraph.

\paragraph{Interception all messages}

In Marea, not being able to intercept messages is a problem because those messages will be directly executed by the proxy instead of being intercepted. Therefore, for Marea it is necessary that Ghost is able to intercept all messages.

\subsubsection{Marea results}

Thanks to all the mentioned advantages of Ghost, among other reasons, Marea is able to significantly reduce the memory used by applications.
Benchmarks (deeply described in~\cite{Mart12d}) demonstrate that the memory footprint of different representative real-world applications can be reduced from 25\% to 40\%.
Marea is not the focus of this paper but basically, these benchmarks compare memory consumption of real applications loaded in a Pharo environment ready for production (already shrinked) with and without Marea.
One of the reasons of Marea's good results is because Ghost proxies use a small memory footprint.

\subsection{Method Wrappers} Method wrappers \cite{Bran98a} control the execution of methods by wrapping them into other objects (\ie the wrappers) that perform some task on method evaluation. The implementation of method wrappers is straightforward with Ghost. In fact, it has been already shown with the \ct{self log:} in the example of \ct{SimpleForwarderHandler} of \secref{proxyMethods}. We can use the regular \ct{TargetBasedProxy} class with a \ct{CompiledMethod} as a target and implement the following handler:

\begin{code}{}
AbstractMethodWrapper >> handleInterception: anInterception
       | answer |
       self preExecutionFor: anInterception.
       answer := anInterception message sendTo: anInterception proxy proxyTarget.
       self postExecutionFor: anInterception.
       ^ answer
\end{code}

Then concrete subclasses implement \ct{preExecutionFor:} and \ct{postExecutionFor:}. In addition, if the user needs to wrap all methods of a class, a better approach is to directly create a \ct{TargetBasedClassProxy} rather than creating a \ct{TargetBasedProxy} for every method.

\sd{add some benchmarks.}

\section{Related Work}
\seclabel{relatedWorks}

\mmp{maybe we can shorten this section?}

\noury{We need to compare reflective facilities of other Smalltalk dialects. For example, IIRC VW does not support methodDict set to nil.}
\subsection{Proxies in dynamic languages}
\seclabel{proxysDinamicLanguages}


\paragraph{Objective-C}
Objective-C provides a proxy implementation called \ct{NSProxy} \footnote{Apple developer library documentation: \url{http://developer.apple.com/library/ios/\#documentation/cocoa/reference/foundation/Classes/NSProxy_Class/Reference/Reference.html}.}. This solution consists of an abstract class \ct{NSProxy} that implements the minimum number of methods needed to be a root class. Indeed, this class is not a subclass of \ct{NSObject} (the root class of the hierarchy chain), but a separate root class (like subclassing from \ct{nil} in Smalltalk). The intention is to reduce method conflicts between the proxified object and the proxy. Subclasses of \ct{NSProxy} can be used to implement distributed messaging, future objects or other proxies usage. Typically, a message to a proxy is forwarded to a proxified object which can be an instance variable in a \ct{NSProxy} subclass.

Since Objective-C is a dynamic language, it needs to provide a mechanism like the Smalltalk \ct{doesNotUnderstand:} when an object receives a message that cannot understand. When a message is not understood, the Objective-C runtime sends \ct{methodSignatureForSelector:} to see what kind of argument and return types are present. If a method signature is returned, the runtime creates a \ct{NSInvocation} object describing the message being sent and then sends \ct{forwardInvocation:} to the object. If no method signature is found, the runtime sends \ct{doesNotRecognizeSelector:}.

\ct{NSProxy} subclasses must override the \ct{forwardInvocation:} and \ct{methodSignatureForSelector:} methods to handle messages that they do not  implement themselves. By implementing the method \ct{forwardInvocation:}, a subclass can define how to process the invocation \eg forwarding it over the network. The method \ct{methodSignatureForSelector:} is required to provide argument type information for a given message. A subclass' implementation should be able to determine the argument types for the messages it needs to forward and it should be able to build a \ct{NSMethodSignature} object accordingly. Note that, from this point of view, Objective-C is not so dynamic.

To sum up, the developer needs to subclass \ct{NSProxy} and implement the \ct{forwardInvocation:} to handle messages that are not understood by itself. One of the drawbacks of this solution is that the developer does not have control over the methods that are implemented in \ct{NSProxy}. For example, such class implements the methods \ct{isEqual:}, \ct{hash}, \ct{class}, etc. This is a problem because those messages will be understood by the proxy instead of being intercepted. This solution is similar to the common solution in Smalltalk with \ct{doesNotUnderstand:}.

\paragraph{Ruby}
In Ruby, there is a proxy implementation which is called \ct{Delegator}. It is just a class included in Ruby standard library but which can be easily modified or implemented from scratch. Similar to Objective-C and Smalltalk (and indeed, to most dynamic languages), Ruby provides a mechanism used when an object receives a message that it does not understand. This method is called \ct{method\_missing(aSelector, *args)}. Moreover, from Ruby version 1.9, there is a new minimal class called \ct{BasicObject} which understands a few methods and is similar to \ct{ProtoObject} in Pharo.

Ruby's proxies are similar to Smalltalk's proxies using \ct{doesNotUnderstand:} and to Objective-C' \ct{NSProxy} as they have a minimal object (subclass from \ct{BasicObject}) and implement \ct{method\_missing(aSelector, *args)} to intercept messages.


\paragraph{Javascript}

Mozilla's Spidermonkey JavaScript engine has long included a nonstandard way of intercepting method calls based on Smalltalk's \ct{doesNotUnderstand:}. The equivalent method is named \ct{noSuchMethod}. Such solution is not stratified and it only intercepts the messages that are not understood.

Van Cutsem et al. implemented what they call ``Dynamic Proxies'' \cite{Vanc10a} which are objects that act like normal objects but whose behavior is controlled by another object known as handler. They provide a  clear division between proxies and handlers. Similarly to our possibility of creating proxies for methods, they can create proxies for Javascript functions since they are objects too. As well as we do, they have several reasons to avoid intercepting \ct{===} (object identity in Javascript).

From what we can understand, their solution requires changes in the VM. Contrary to Smalltalk where almost everything is a message send, in Javascript, apart from function calls, there are operators. We understand that the VM needs to be changed so that each of these operators can check whether the receiver is a proxy or not redirecting the invocation to the handler rather than following the normal steps when the answer is positive. They also provide way for the user to specify a list of properties that are answered directly by the proxy instead of being intercepted.

The authors also claim that having only one trap message \eg \ct{doesNotUnderstand:} does not scale if we were to introduce additional traps to intercept not only method invocation, but also property access, assignment, lookup, enumeration, etc. Nonetheless, some items of such list do not apply to all programming languages. For example, in Pharo, the only way to access instance variables from outside an object is by sending a message so it is a message send. Enumerations are also just messages. The lookup can be intercepted with Ghost by using proxies for classes. So, from such list, the only item Ghost is unable to intercept is  assignment.

Javascript is a prototype-based language which, besides function calls, it has more language constructs. Smalltalk is an object-oriented programming language and system where most of the language constructions are message sends. Furthermore, their work takes a more reflective standpoint. They want a way to reify every operation such as message send, instance variable access, etc. In Smalltalk, this is not needed for proxies. Nevertheless, if we want to do behavioral reflection and change \eg instance variable access (to make it persistent, for example), then we would need a way to intercept them even when accessing from within the object.


\subsection{Proxies in static languages}
\seclabel{proxysStaticLanguages}

\paragraph{Java}
Java, being a statically-typed language, supports quite limited proxies called \emph{Dynamic Proxy Classes}.
It relies on the \ct{Proxy} class from the {java.lang.reflect} package.

\begin{aquote}{Java Dynamic Proxies. The Java Platform 1.5 API Specification}
``Proxy provides static methods for creating dynamic proxy classes and instances, and it is also the superclass of all dynamic proxy classes created by those methods.

Each proxy instance has an associated invocation handler object, which implements the interface \ct{InvocationHandler}. A method invocation on a proxy instance through one of its proxy interfaces will be dispatched to the invoke method of the instance's invocation handler, passing the proxy instance, a \ct{java.lang.reflect.Method} object identifying the method that was invoked, and an array of type \ct{Object} containing the arguments. The invocation handler processes the encoded method invocation as appropriate and the result that it returns will be returned as the result of the method invocation on the proxy instance.''
\end{aquote}

The creation of a dynamic proxy class can only be done by providing a list of java interfaces that should be implemented by the generated class.
All messages corresponding to the declarations in the provided interfaces will be intercepted by a proxy instance of the generated class and delegated to a handler object.

Java proxies have the following limitations:

\begin{itemize}
\item The user \emph{cannot} create a proxy for instances of a class which has not all its methods declared in interfaces. This means that, if the user wants to create a proxy for a domain class, he is forced to create an interface for it. Eugster \cite{Eugs06a} proposed a solution which provides proxies for classes. There is also a third-party framework based on bytecode manipulation called CGLib \footnote{cglib Code Generation Library: \url{http://cglib.sourceforge.net}.} which provides proxies for classes.
\item \emph{Only} the methods defined in the interface will be intercepted which is a big limitation.\noury{Explain a bit more. I think that this concerns two things: inherited methods + methods declared in classes and not methods.}
\item Java interfaces do not support private methods. Since Java proxies require interfaces, private methods cannot be intercepted either. Depending on the proxy usage, this can be a problem.
\item Proxies are subclass from \ct{Object} forcing them to understand several messages \eg ~\ct{getClass}. So, a proxy answers its own class instead of the target's one. Therefore, the proxy is not transparent and it is not fully stratified. Moreover, there are some specific exceptions: when the messages \ct{hashCode}, \ct{equals} or  \ct{toString} (declared in \ct{Object}) are sent to a proxy instance, they are encoded and dispatched to the invocation handler's invoke method, \ie they are intercepted.
\end{itemize}

\paragraph{.Net}
Microsoft's .NET platform \cite{microsoftNET} proposes a closely related concept of Java dynamic proxies with nearly the same limitations. There are other third-party libraries like \emph{Castle DynamicProxy}\footnote{Castle DynamicProxy library: \url{http://www.castleproject.org/dynamicproxy/index.html}} or \emph{LinFu}\footnote{Linfu proxies framework: \url{http://www.codeproject.com/KB/cs/LinFuPart1.aspx}}. DynamicProxy differs from the proxy implementation built into .NET which requires the proxified class to extend \ct{MarshalByRefObject}.
This is a too heavy constraint since instances of classes that do not subclass \ct{MashalByRefObject} cannot be proxified.
\mmp{are we sure about this?}
In LinFu, every generated proxy dynamically overrides all of its parent's virtual methods. Each of its respective overridden method implementations delegates each method call to the attached interceptor object. However, none of them can intercept non-virtual methods.

\subsection{Comparison}
\seclabel{comparison}

Statically typed languages, such as Java or .NET, support quite limited proxies \cite{Barne03}.
Java and .Net suffer from the lack of replacement issue and transparency. \noury{explain transparency}
\mmp{but it was already explained in the paper...if I need to re-explain everything in the related work it would be long and full of repetitions}
Another problem in Java is that one cannot build a proxy with fields storing any specific data. Therefore, one has to put everything in the handler meaning no handler sharing is possible which ends in a bigger memory footprint. Proxies are far more powerful, flexible, transparent and easy to implement in dynamic languages than in static ones.

Dynamic languages just need two features to implement a basic Proxy solution: 1) a mechanism to handle messages that are not understood by the receiver object and 2) a minimal object that understands a few or no messages so that the rest are managed by the mentioned mechanism.  Objective-C NSProxy, Ruby Decorator, etc, all work that way. Nevertheless, none of them solve all the problems mentioned in this paper:

\begin{description}

\item [ Uniformity. ]  All the investigated solutions create proxies for specific objects but none of them are able to create proxies for classes or methods.

\item[ Object replacement.]  Some proxy solutions can create a proxy for a particular object \ct{X}. The user can then use that proxy instead of the original object. The problem is that there may be other objects in the system referencing \ct{X}. Without object replacement, those references will still be pointing to \ct{X} instead of pointing to the proxy. Depending on the proxies usage, this can be a limitation.

\item[ Memory footprint.]  None of the solutions take special care of the memory usage of proxies. This is a real limitation when proxies are being used to save memory.

\end{description}

\section{Conclusion}
\seclabel{conclusion}

In this paper, we described the need for proxies, their different usages and common problems while trying to implement them.  We introduced Ghost: a generic proxy implementation on top of Pharo Smalltalk.

Our solution provides uniform proxies not only for regular instances, but also for classes and methods. Ghost optionally supports object replacement. In addition, Ghost proxies can have a small memory footprint. Proxies are powerful, easy to use and extend and its overhead is low.
\mmp{I should really check the overhead. Check mail with subject ``Overhead of proxies''}

Ghost's implementation takes advantages of Pharo VM reflective facilities and hooks. Nevertheless, we believe that such specific features, provided by Pharo and its VM, can also be ported to other dynamic programming language.




\bibliographystyle{alpha}
\bibliography{rmod,others,extras}

%

\end{document}